# Exact solution of two-dimensional (2D) Ising model with a transverse field: a low-dimensional quantum spin system

Zhidong Zhang

Shenyang National Laboratory for Materials Science, Institute of Metal Research, Chinese Academy of Sciences, 72 Wenhua Road, Shenyang, 110016, P.R. China

The exact solution of ferromagnetic two-dimensional (2D) Ising model with a transverse field, which can be used to describe the critical phenomena in low-dimensional quantum spin systems, is derived by equivalence between the ferromagnetic 2D Ising model with a transverse field and the ferromagnetic three-dimensional (3D) Ising model. The results obtained in this work can be extended to be suitable for the antiferromagnetic 2D Ising model with a transverse field in the cases without frustration.

The corresponding author: Z.D. Zhang, Tel: 86-24-23971859, Fax: 86-24-23891320, e-mail address: zdzhang@imr.ac.cn

Recently, low-dimensional systems have attracted intensive interest. The critical phenomena exist widely in various physical systems, which have been one of central topics in physics, for bulk systems as well as low-dimensional systems. Among the physical models describing the critical phenomena, the Ising model represents a class of universality [1-6]. Critical phenomena in classical systems with a phase transition at/near a critical temperature, whose Hamiltonians are described by scalar variables such as Ising spins, have been studied intensively since Onsager found the exact solution of the two-dimensional (2D) Ising model [2]. The exact solution of three-dimensional (3D) Ising model is a long-standing well-known problem in physics. Parallelly, critical phenomena at/near quantum phase transitions in quantum spin systems have also attracted great interest of physics community [7-12]. A quantum system can undergo a continuous phase transition at the absolute zero of temperature as some parameter entering its Hamiltonian is varied [7]. This parameter can be the magnetic field in a quantum Hall sample (which controls the transition between quantized Hall plateaus), the pressure applied on a heavy fermion compound [9], the charging energy in Josephson-junction arrays (which controls their superconductor-insulator transition), doping in the parent compound of a high-Tc superconductor (which destroys the antiferromagnetic spin order) [10], disorder in a conductor near its metal-insulator transition (which determines the conductivity at zero temperature), and so on [7]. The Ising model with a transverse field can be utilized to describe the quantum phase transitions and the critical phenomena in the quantum spin systems. The one-dimensional (1D) Ising model with a transverse field is solved exactly by transforming the set of Pauli operators to a new set of Fermi operators [11]. For the transverse-field Ising model, the quantum–classical mapping can be easily demonstrated at a microscopic level [8]. A single quantum spin can be mapped onto a classical Ising chain and such a mapping can be generalized to a mapping between a d-dimensional transverse-field (quantum) Ising model and a (d + 1)-dimensional classical Ising model [8]. Suzuki [12] proved rigorously equivalence between the d-dimensional quantal spin systems and the (d+ 1)-dimensional classical Ising systems, regarding the ground state, critical exponents and spin correlation. Up to date, no exact solution for the 2D Ising model with a transverse field has been reported yet. However, the exact solution of the 3D Ising model is very important for understanding other many-body interacting spin systems. In the previous work [3-5], I proposed the two conjectures in [3] and then proved them in collaboration with Suzuki and March [5] for solving the exact solution of the ferromagnetic 3D Ising model with all the interactions positive. The exact solution of the ferromagnetic 3D Ising model provides an opportunity to determine the exact solution of the ferromagnetic 2D Ising model with a transverse field. In this work, I shall prove seven theorems for ground state, partition function, specific heat, spontaneous magnetization, spin correlation, susceptibility, critical exponents of the ferromagnetic 2D Ising model with a transverse field. This is accomplished by simply employing the equivalence between the ferromagnetic 2D Ising model with a transverse field and the ferromagnetic 3D Ising model. However, the results obtained in this work can be applied directly to derive the exact solution for the antiferromagnetic 2D Ising model with a transverse field in the cases without frustration.

Theorem 1: The ground state of the ferromagnetic 2D Ising model with a transverse field is equivalent to the ferromagnetic 3D Ising model.

Theorem 2: The partition function Z of the ferromagnetic 2D Ising model with a transverse field is equivalent to the ferromagnetic 3D Ising model for the lattice size N =nml, which can be represented as:

$$N^{-1} \ln Z = \ln 2$$
$$+ \frac{1}{2(2\pi)^4} \int_{-\pi}^{\pi} \int_{-\pi}^{\pi} \int_{-\pi}^{\pi} \int_{-\pi}^{\pi} \ln \Big[ \cosh 2 K_1 \cosh 2 (K_2 + K_3 + K_4)$$
$$- \sinh 2 K_1 \cos \omega'$$
$$- \sinh 2 (K_2 + K_3 + K_4)(|w_x| \cos \phi_x \cos \omega_x + |w_y| \cos \phi_y \cos \omega_y$$
$$+ |w_z| \cos \phi_z \cos \omega_z) \Big]$$
$$d\omega' d\omega_x d\omega_y d\omega_z$$

with variables $K_1 = J_1/H_T$, $K_2 = J_2/H_T$, $K_3 = (1/2)\log(\coth(1/q))$, $K_4 = (1/2)\log(\coth(1/q))$ $J_2/J_1$. $|w_x| = |w_y| = |w_z| = 1$. The topological phases $\phi_x$, $\phi_y$, and $\phi_z$ at finite $H_T$ are equal to $2\pi$, $\pi/2$ and $\pi/2$, respectively.

Theorem 3: The specific heat of the ferromagnetic 2D Ising model with a transverse field is equivalent to the ferromagnetic 3D Ising model, as derived in [3].

Theorem 4: The spontaneous magnetization of the ferromagnetic 2D Ising model with a transverse field is equivalent to the ferromagnetic 3D Ising model, which can be represented as:

$$M = \left[ 1 - \frac{16 x_1^2 x_2^2 x_3^2 x_4^2}{(1-x_1^2)^2 (1-x_2^2 x_3^2 x_4^2)^2} \right]^{\frac{3}{8}}$$

with parameters $x_i = e^{-2K_i}$ (i = 1, 2, 3, 4).

Theorem 5: The spin correlation (including the correction length and correlation function at the critical point) of the ferromagnetic 2D Ising model with a transverse field are equivalent to the ferromagnetic 3D Ising model, as derived in [3].

Theorem 6: The susceptibility of the ferromagnetic 2D Ising model with a transverse field is equivalent to the ferromagnetic 3D Ising model, as derived in [3].

Theorem 7: The static critical exponents of the ferromagnetic 2D Ising model with a transverse field are equivalent to the ferromagnetic 3D Ising model, which are determined to be $\alpha = 0$, $\beta = 3/8$, $\gamma = 5/4$, $\delta = 13/3$, $\eta = 1/8$ and $\nu = 2/3$, satisfying the scaling laws.

**Proof of Theorems 1-7:**

The 2D Ising model with a transverse field is described by the Hamiltonian [12]:

$$H = -\sum_{i,j \in R^d}^{nm} J_{ij} \sigma_i^z \sigma_j^z - \mu_0 H \sum_{i,j \in R^d}^{nm} \sigma_{i,j}^z - H_T \sum_{i,j \in R^d}^{nm} \sigma_{i,j}^x$$

in which it consists of three terms: interactions $J_{ij}$ between spins located at lattice points $R^d$ in a plane (d = 2) with the total lattice points nm, a magnetic field H term and a transverse field term $H_T$. It is noticed that in order to be consistent with the notations in our previous work [3-5], we use (i, j, k) running from (1, 1, 1) to (n, m, l) to denote the 3D lattice with the lattice size N = nml, and use the parameters p and q to replace the parameter n and m defined in Suzuki's paper [12], thus here the total lattice point along the third dimension is l = pq. For simplicity, we consider only the ferromagnetic interactions $J_1 > 0$ and $J_2 > 0$ between the nearest neighboring spins along two crystallographic axes in the plane and, the case at absence of the magnetic field (H = 0). Following the Suzuki' procedure [12], we introduce a dimensionless parameter p = $H_T/(k_B T)$. In the limit T → 0, we have T = $H_T/(k_B p)$. Suzuki [12] proved that the transverse field term in a d-dimensional Ising model with a transverse field can be transformed into an interaction term along the additional dimension in a (d+1)-dimensional Ising model (see Eqs. (4.3) and (4.4) in his paper [12]). Such a transformation reduces to an effective Hamiltonian written as [12]:

$$H_{eff}^{3D} = \frac{1}{qH_T}\sum_{i,j \in R^d}^{nm}\sum_{k=1}^{l} J_{i,j}\sigma_{i,k}^z \sigma_{j,k}^z + \frac{1}{2}\left(log\left(coth\left(\frac{1}{q}\right)\right)\right)\sum_{i,j \in R^d}^{nm}\sum_{k=1}^{l} \sigma_{i,j,k}^z \sigma_{i,j,k+1}^z$$

As mentioned above, here $R^d$ with d = 2 and, the 3D lattice points (with N = nml) are described by (i, j, k) running from (1, 1, 1) to (n, m, l). Of course, it is true for the equivalence between the ferromagnetic 2D Ising model with a transverse field and the ferromagnetic 3D Ising model. Suzuki [12] proved that the ground state energy of the Ising model in d dimensions can be expressed by the partition function of the effective (d+1)-dimensional Ising model, in which $H_T$ plays the role of temperature. Therefore, if $H_T > H_{T,c}$ (critical point), the system is paramagnetic and if $H_T < H_{T,c}$, the system becomes ferromagnetic in the z-direction. At/near the critical point of the quantum phase transition, there exist the critical phenomena and we can observe the scaling behaviors of the physical properties with the critical exponents. It is worth noting that such a correspondence is valid for all finite values of q [12]. The limit p → ∞ can be taken for q fixed [12], which means that the thermodynamic limit can reach also along the third dimension as p → ∞, and l = pq → ∞.

After the ferromagnetic 2D Ising model with a transverse field is transformed into the ferromagnetic 3D Ising model, we have three parameters for interactions along the dimensions: $K_1 = J_1/H_T$, $K_2 = J_2/H_T$, $K_3 = (1/2)log(coth(1/q))$, corresponding to those (K = $J_1/(k_BT)$, K' = $J_2/(k_BT)$, K" = $J_3/(k_BT)$) in the ferromagnetic 3D Ising model we studied in [3-5]. Clearly, after such a transformation, the two parameters $J_1/H_T$ and $J_2/H_T$ representing the interactions in the plane are defined with the transverse field $H_T$, which can be adjusted by the transverse field $H_T$. But the third parameter $(1/2)\log(\coth(1/q))$ for the third dimension is not related with the transverse field $H_T$, thus does not change with varying $H_T$. Note that the total lattice point $l = pq$ along the third dimension is related with the transverse field $H_T$ and the temperature T (see the above definition $p = H_T/(k_BT)$). However, in the ferromagnetic 3D Ising model we studied before [3-5], all the three parameters K, K' and K'' for interactions along three dimensions are defined with temperature T, changing with varying temperature. This is the only difference between the ferromagnetic 3D Ising model transformed from the ferromagnetic 2D Ising model with a transverse field and the ferromagnetic 3D Ising model we studied previously in [3-5] for classical systems. Keeping in mind such a difference, we can employ directly our previous results obtained in [3-5] to investigate the physical properties of the ferromagnetic 2D Ising model with a transverse field for quantum systems, by a direct replacement of these parameters.

I proposed the two conjectures in [3] for solving the exact solution of the ferromagnetic 3D Ising model, which are repeated below.

**Conjecture 1 [3]:** The topologic problem of a 3D Ising system can be solved by introducing an additional rotation in a four-dimensional (4D) space, since the knots in a 3D space can be opened by a rotation in a 4D space. One can find a spin representation in $2^{n \cdot l \cdot o}$-space for this additional rotation in $2n \cdot l \cdot o$-space with $o = (n \cdot l)^{1/2}$. Meanwhile, the matrices **V₁, V₂** and **V₃** have to be represented and rearranged, also in the $2n \cdot l \cdot o$-space.

**Conjecture 2 [3]:** The weight factors w_x, w_y and w_z, varying in range of [-1, 1], on the eigenvectors represent the contribution of $e^{i\frac{t_x\pi}{n}}$, $e^{i\frac{t_y\pi}{l}}$ and $e^{i\frac{t_z\pi}{o}}$ in the 4D space to the energy spectrum of the system.

In collaboration with Suzuki and March [5], I then proved the four theorems as follows:

**Trace Invariance Theorem [5]:** The partition function of the 3D Ising model is unchanged by adding k terms of unit matrices in the direct product of the original transfer matrices and by compensation of a factor of $2^k$, and also by adjusting the sequence of the unit matrices with other matrices to separate the exponential factors for different row in the transfer matrices to the sub-transfer matrices.

**Linearization Theorem [5]:** A linearization process can be performed on the sub-transfer matrices of the 3D Ising model for each row, to linearize the non-linear terms.

**Local Transformation Theorem [5]:** A local transformation can be performed on the sub-transfer matrices of the 3D Ising model for each row, which trivializes the non-trivial topological structure and generates the topological phases on the eigenvectors.

**Commutation Theorem [5]:** The non-commutative behaviors of the operators in the linearization and local transformation processes of the 3D Ising model can be dealt with to be commutative in the framework of the Jordan-von Neumann-Wigner procedure with Jordan algebras, by performing a time average of t systems of the 3D Ising models.

These four theorems [5] give a positive answer to my two conjectures [3], which means that the exact solution based on the conjectures is correct. Therefore, we can utilize the exact solution of the ferromagnetic 3D Ising model obtained in [3] to derive the exact solution of the ferromagnetic 2D Ising model with a transverse field, by employing simply the equivalence between the two models as proved by Suzuki [12] and the relations obtained above for the parameters. The ground state, the partition function, the specific heat, the spontaneous magnetization, the spin correlation, the susceptibility as well as the static critical exponents of the ferromagnetic 2D Ising model with a transverse field are equivalent to those of the ferromagnetic 3D Ising model, which are explicitly derived in [3-5]. Note that this work does not focus on the critical exponents of the dynamic process of the quantum spin systems. In order to deal with the non-trivial topological structure in the ferromagnetic 3D Ising model, we have introduced a rotation with an angle K''' = K'K''/K in an additional dimension for the local transformation [3-5]. For the ferromagnetic 2D Ising model with a transverse field, we have to do the same process now with an angle $K_4 = K_2 K_3/K_1 = (1/2)\log(\coth(1/q))$ $J_2/J_1$. The weight factors $w_x$, $w_y$ and $w_z$ in the eigenvectors in Eq. (33) of ref. [3] have been generalized as complex numbers $|w_x| e^{i\phi_x}$, $|w_y| e^{i\phi_y}$, and $|w_z| e^{i\phi_z}$ with phases $\phi_x$, $\phi_y$, and $\phi_z$ [4,13]. $|w_x| = |w_y| = |w_z| = 1$. Thus, the weight factors $w_x$, $w_y$ and $w_z$ in the partition function (Eq. (49) of ref. [3]) are replaced by $|w_x| \cos \phi_x$, $|w_y| \cos \phi_y$ and $|w_z| \cos \phi_z$, respectively [4,13], since only the real part of the phase factors appears in the eigenvalues. The topological phases $\phi_x$, $\phi_y$, and $\phi_z$ at finite $H_T$ (corresponding to finite temperature in the 3D Ising model [5]) are determined to equal to $2\pi$, $\pi/2$ and $\pi/2$, respectively [5]. This completes the proof of Theorems 1-7.

□

**Remark:**

Because the exact solution for the antiferromagnetic 3D Ising model with all the negative interactions but without frustration has the same formula as the exact solution obtained in [3-5] for the ferromagnetic 3D Ising model with all the interactions positive, the results obtained in this work can be extended to be suitable for the antiferromagnetic 2D Ising model with a transverse field in the cases without frustration. The theoretical model proposed in this paper is in the same universality class with the 3D Ising model at a zero magnetic field. The adjusting parameter in the present model for quantum phase transitions can be a transverse magnetic field, a transverse electric field, a pressure, a charging energy, a doping, or a disorder degree, etc., while the materials can be different 2D quantum systems with Ising spins, such as magnets, ferroelectric materials, quantum Hall systems, heavy fermions, Josephson junctions, superconductors, etc. [7-12,14-16]. In a recent work [17], the exact solution of the 3D $Z_2$ lattice gauge theory is derived by the duality between the 3D $Z_2$ lattice gauge theory and the 3D Ising model. Note that the 3D $Z_2$ lattice gauge theory can be mapped also to the ferromagnetic 2D Ising model with a transverse field [18].

**Acknowledgements**

This work has been supported by the National Natural Science Foundation of China under grant number 52031014, by the State key Project of Research and Development of China (No.2017YFA0206302).